\documentclass[english]{IEEEtran}
\PassOptionsToPackage{ruled,linesnumbered}{algorithm2e}
\usepackage[T1]{fontenc}
\usepackage[latin9]{inputenc}
\usepackage{verbatim}
\usepackage{refstyle}
\usepackage{units}
\usepackage{algorithm2e}
\usepackage{amsmath}
\usepackage{amsthm}
\usepackage{amssymb}
\usepackage{graphicx}

\makeatletter

\AtBeginDocument{\providecommand\thmref[1]{\ref{thm:#1}}}
\AtBeginDocument{}
\AtBeginDocument{\providecommand\algref[1]{\ref{alg:#1}}}
\AtBeginDocument{\providecommand\figref[1]{\ref{fig:#1}}}
\RS@ifundefined{subsecref}
{\newref{subsec}{name = \RSsectxt}}
{}
\RS@ifundefined{thmref}
{\def\RSthmtxt{theorem~}\newref{thm}{name = \RSthmtxt}}
{}
\RS@ifundefined{lemref}
{\def\RSlemtxt{lemma~}\newref{lem}{name = \RSlemtxt}}
{}

\theoremstyle{plain}
\newtheorem{thm}{\protect\theoremname}
\theoremstyle{remark}

\newcommand\ztag[1]{%
	\def\@currentlabel{#1}%
	\gdef\tmp{%
		\addtocounter{equation}{-1}%
		\def\theequation{#1}}%
	\aftergroup\aftergroup\aftergroup\aftergroup\aftergroup\aftergroup
	\aftergroup\aftergroup\aftergroup\aftergroup\aftergroup\aftergroup
	\aftergroup\aftergroup\aftergroup\aftergroup\aftergroup\aftergroup
	\aftergroup\aftergroup\aftergroup\aftergroup\aftergroup\aftergroup
	\aftergroup\aftergroup\aftergroup\aftergroup\aftergroup\aftergroup
	\aftergroup
	\tmp}

\usepackage{mathtools,empheq}
\usepackage{cite}
\usepackage{caption}
\usepackage{refstyle}
\usepackage[subrefformat=parens,labelformat=parens]{subfig}
\usepackage[esvectg]{overarrows}
\usepackage{pgfplots}
\usepackage{grffile}
\pgfplotsset{compat=newest}
\usetikzlibrary{plotmarks}
\usepgfplotslibrary{patchplots}
\usepackage{amsmath}
\usepackage[T1]{fontenc}
\usepackage[latin9]{inputenc}
\usepackage{pgfplots}
\usepackage{grffile}
\pgfplotsset{compat=newest}
\usetikzlibrary{plotmarks}
\usepgfplotslibrary{patchplots}
\usepackage{amsmath}
\usepackage{xcolor}
\usepackage{tikz}
\usepackage{nicefrac, xfrac}
\usepackage{cases}

\newcommand{\herm}{^{\mathsf{H}}}
\newcommand{\TX}{{\mathtt{T}}}
\newcommand{\RX}{{\mathtt{R}}}

\newcommand{\trans}{^{\mathsf{T}}}
\usepackage{amsmath}
\DeclareMathOperator{\Tr}{Tr}
\DeclareMathOperator{\st}{s.t.}
\DeclareMathOperator{\diag}{diag}

\DeclareMathOperator{\vect}{vec}
\DeclareMathOperator{\vecd}{vec_{d}}

\allowdisplaybreaks[1]

\newref{fig}{refcmd = Fig.~\ref{#1}}
\newref{subfig}{refcmd = Fig.~\subref*{#1}}
\newref{alg}{refcmd = \textbf{Algorithm~\ref{#1}}}
\newref{app}{refcmd = Appendix~\ref{#1}}

\newref{subsec}{refcmd = Section~\ref{#1}}
\newref{sec}{refcmd = Section~\ref{#1}}
\renewcommand*{\thmref}[1]{\textbf{Theorem~\ref{thm:#1}}}

\AtBeginDocument{\providecommand\eqref[1]{(\ref{eq:#1})}}

\makeatother

\usepackage{babel}
\providecommand{\remarkname}{Remark}
\providecommand{\theoremname}{Theorem}

\begin{document}
	\title{{ Scaling Achievable Rates in SIM-aided MIMO Systems with Metasurface Layers: A Hybrid Optimization Framework}}
	\author{Eduard E. Bahingayi,\IEEEmembership{Member, IEEE}, Nemanja~Stefan~Perovi\'c, \IEEEmembership{Member, IEEE},
		and Le-Nam Tran, \IEEEmembership{Senior Member, IEEE}\thanks{The authors are with the School of Electrical and Electronic Engineering,  University College Dublin, Belfield, Dublin 4, D04 V1W8 Ireland. Email: \{eduard.bahingayi, nemanja-stefan.perovic, nam.tran\}@ucd.ie}}
	\maketitle
	\begin{abstract}
		We investigate the achievable rate (AR) of a stacked intelligent metasurface (SIM)-aided holographic multiple-input multiple-output (HMIMO) system by jointly optimizing the SIM phase shifts and power allocation. Contrary to earlier studies suggesting that the AR decreases when the number of metasurface layers increases past a certain point for \emph{a fixed SIM thickness}, our findings demonstrate a consistent increase. To achieve this, we introduce two problem formulations: one based on directly maximizing the AR (RMax) and the other focused on minimizing inter-stream interference (IMin). To solve the RMax problem, we apply Riemannian manifold optimization (RMO) and weighted minimum mean square error (WMMSE) methods to optimize the SIM phase shifts and power allocation alternately. For the IMin problem, we derive an efficient algorithm that iteratively updates each meta-atom's phase shift using a closed-form expression while keeping others fixed. Our key contribution is a hybrid optimization framework, where the IMin solution initializes the SIM phase shifts in the first algorithm. This hybrid strategy enhances AR performance across varying numbers of metasurface layers. Simulation results demonstrate that the proposed algorithms outperform existing benchmarks. Most importantly, we show that increasing the number of metasurface layers while keeping the SIM thickness fixed leads to significant AR improvements.
	\end{abstract}
	

	\section{INTRODUCTION}
	
	The stacked intelligent metasurface (SIM)-aided holographic multiple-input multiple-output (HMIMO) system is a revolutionary approach for wireless communications, aiming at improving spectral and energy efficiency \cite{an2023stacked}. This novel technology incorporates multiple passive metasurface layers into the transceiver architecture, each consisting of numerous meta-atoms. The signals propagate through meta-atoms across layers, each meta-atom acting as a secondary signal source for the next layer. Unlike metallic antennas, meta-atoms require low cost and low power consumption. Moreover, in SIM-aided HMIMO systems, signal precoding and combining occur directly in the native electromagnetic (EM) wave domain by intelligently controlling the phase shifts of the meta-atoms. This approach reduces the need for complex digital baseband processing typically required in conventional MIMO systems \cite{an2023stacked,an2024stacked}. Therefore, SIM-aided HMIMO systems are expected to provide improved energy and spectral efficiency with minimal additional hardware complexity \cite{perovic2024energy,an2024stacked}.
	
	Recent studies have opened up new avenues for future research by exploring the performance of SIM-based systems across various metrics. A brief overview of the existing literature is in order. In \cite{an2023stacked}, closely related to our work, the authors investigated the achievable rate (AR) of SIM-aided HMIMO systems by solving a channel fitting optimization problem using the projected gradient (PG) method to optimize both transmit and receive SIM phase shifts. This study was extended in \cite{papazafeiropoulos2024achievable} by incorporating digital precoding and combining. The work in \cite{NemanjaSIM} studied the mutual information maximization problem for SIM-aided HMIMO systems by utilizing the cutoff rate as an alternative metric. Meanwhile, the authors in \cite{perovic2024energy} studied the energy efficiency of SIM-based systems. Other use cases of SIM-aided wireless communication systems include multi-user MIMO \cite{lin2024stacked}, MIMO integrated sensing and communication \cite{niu2024stacked}, and Cell-Free Massive MIMO \cite{hu2024joint}.
	
	In this work, we aim to further explore the AR of SIM-aided HMIMO systems by jointly optimizing the SIM phase shifts and power allocation. To this end, we formulate two optimization problems: the RMax problem, which focuses on directly maximizing the AR, and the IMin problem, which aims to minimize inter-stream interference. To solve the RMax problem, we employ a Riemannian manifold optimization (RMO) method for phase shift optimization and the weighted minimum mean square error (WMMSE) method to optimize power allocation in the alternating optimization (AO) manner. The benefits of considering the IMin formulation are twofold. First, it leads to a low-complexity algorithm, where a closed-form solution can be derived to iteratively optimize each meta-atom phase shift in the SIM layers with others fixed. Second, it can significantly suppress inter-stream interference, especially when the number of SIM layers is high, thereby enabling the efficient application of the water-filling (WF) algorithm to find a near-optimal power allocation policy.
    
    Importantly, our formulations challenge the existing belief that the AR of SIM-aided HMIMO systems degrades when the number of SIM layers exceeds a certain point for a fixed SIM thickness, a conclusion reported in several early studies \cite{an2023stacked,an2024stacked,papazafeiropoulos2024achievable,lin2024stacked}. Through numerical experiments, we find that this degradation arises solely \emph{not} from the physical limitations of the SIMs but from the inefficacy of the iterative optimization methods used in these studies. Specifically, iterative methods, such as those proposed in \cite{an2023stacked} and \cite{papazafeiropoulos2024achievable}, are highly sensitive to the initial points and often converge to suboptimal solutions in large-scale non-convex problems, particularly when the number of SIM layers increases. In fact, we observe similar issues when applying the first algorithm to solve the RMax problem, where the obtained AR strongly depends on phase shift initialization. To address this critical problem, we propose a hybrid optimization framework that employs the IMin solution to initialize the phase shifts for the RMO method in the first algorithm, leading to significantly improved AR performance across the entire range of metasurface layers.
	
	\emph{Notation}: Upper and lowercase boldface letters denote matrices and vectors, respectively. $x_{i}$ is the $i$-th entry of $\mathbf{x}$, and $\left[\mathbf{X}\right]_{i,j}$ is the $($$i$, $j$$)$-th entry of $\mathbf{X}$. $\mathbf{\left(\cdot\right)}^{*}$, $\mathbf{\left(\cdot\right)}\trans$, and $\mathbf{\left(\cdot\right)}\herm$ denote the conjugate, transpose, and Hermitian, respectively. $\Tr\{\cdot\}$ and $\left\Vert \mathbf{\cdot}\right\Vert $ denote the trace and Euclidean norm. $\diag(\mathbf{\cdot})$ forms a diagonal matrix, while $\vecd(\cdot)$ extracts its diagonal elements as a vector. $\mathbf{\nabla}_{\mathbf{X}}f(\cdot)$ is the gradient of $f$ with respect to (w.r.t) $\mathbf{X}^{\ast}$. $\mathbf{I}_{N}$ is the $N\times N$ identity matrix, and $\otimes$ is the Kronecker product. $\Re\{\cdot\}$, $\arg\{\cdot\}$, and $\bigl|\cdot\bigr|$ denote the real part, angle, and absolute value of a complex number, respectively. $\mathbb{C}$ ($\mathbb{Z}$) stands for the complex (integer) numbers.
	\section{System Model and Problem Formulation\label{sec:SystemModel}}	
	\subsection{System Model}	
	We consider a SIM-aided HMIMO system where a transmitter (TX) sends $S$ data streams to a receiver (RX). As in \cite{an2023stacked,an2024stacked}, digital precoding is not considered; instead, the transmission relies entirely on precoding and combining within the native EM wave domain. In this way, each data stream is handled by a pair of transmit and receive antennas, making the number of antennas at both the TX and RX equal to $S$. The number of metasurface layers at the TX-SIM and RX-SIM are denoted by $L$ and $K$, respectively, with the corresponding layer indices represented as $\mathcal{L}=\{1,\cdots, L\}$ and $\mathcal{K}=\{1,\cdots, K\}$. Additionally, the number of meta-atoms per metasurface layer at the TX-SIM and RX-SIM are $N$ and $M$, respectively, with the corresponding sets of meta-atom indices denoted as $\mathcal{N}=\{1,\cdots, N\}$ and $\mathcal{M}=\{1,\cdots, M\}$.
	
	The propagation coefficient of the $n$-th meta-atom in the $l$-th transmit metasurface layer is represented as $\theta_{\TX,n}^{l}=e^{j\psi_{\TX,n}^{l}}$, where $\psi_{\TX,n}^{l}\in[0,2\pi)$ is the corresponding phase shift. The propagation coefficient vector for the $l$-th transmit metasurface layer is denoted as $\boldsymbol{\theta}_{\TX}^{l}=\left[\theta_{\TX,1}^{l},\cdots,\theta_{\TX,N}^{l}\right]\trans\in\mathbb{C}^{N\times1}$. Similarly, the propagation coefficient of the $m$-th meta-atom in
	the $k$-th receive metasurface layer is given by $\theta_{\RX,1}^{k}=e^{j\psi_{\RX,m}^{k}}$, where $\psi_{\RX,m}^{k}\in[0,2\pi)$ is the corresponding phase shift. The propagation coefficient vector for the $k$-th receive metasurface layer is denoted by $\boldsymbol{\theta}_{\RX}^{k}=\left[\theta_{\RX,1}^{k},\cdots,\theta_{\RX,M}^{k}\right]\trans\in\mathbb{C}^{M\times1}$.
	
	At the TX-SIM, the propagation coefficient matrix between the $(l-1)$-th and $l$-th transmit metasurface layers is denoted by $\mathbf{\Omega}_{\TX}^{l}\in\mathbb{C}^{N\times N},\forall l\in\mathcal{L}/(1)$. Specifically, $[\mathbf{\Omega}_{\TX}^{l}]_{n,n'}$, which represents the signal propagation coefficient between the $n'$-th meta-atom of the $(l-1)$-th layer and the $n$-th meta-atom of the $l$-th layer, is modeled according to Rayleigh?Sommerfeld diffraction theory
	as described in \cite{Lin_2018}:
	\begin{equation}
		[\mathbf{\Omega}_{\TX}^{l}]_{n,n'}=\frac{A\cos\chi_{n,n'}}{d_{n,n'}}(\frac{1}{2\pi d_{n,n'}}-\frac{j}{\lambda})\exp(\frac{j2\pi d_{n,n'}}{\lambda}),\label{eq: Rayleigh-Sommerfield}
	\end{equation}
	where $\lambda$ is the wavelength, $d_{n,n'}$ is the propagation distance between the $n'$-th meta-atom of the $(l-1)$-th layer and the $n$-th meta-atom of the $l$-th layer, $A$ is the surface area of each meta-atom, and $\chi_{n,n'}$ is the angle between the propagation direction and the normal to the $(l-1)$-th transmit metasurface layer. The matrix $\mathbf{\Omega}_{\TX}^{1}\in\mathbb{C}^{N\times S}$ denotes the propagation coefficients between the transmit antenna array and the first transmit metasurface layer, modeled similarly according to (\ref{eq: Rayleigh-Sommerfield}).
	
	Likewise, at the RX-SIM, the propagation coefficient matrix between the $(k-1)$-th and $k$-th receive metasurface layers is denoted by $\mathbf{\Omega}_{\RX}^{k}\in\mathbb{C}^{M\times M},\forall k\in\mathcal{K}/(1)$, whereas $\mathbf{\Omega}_{\RX}^{1}\in\mathbb{C}^{S\times M}$ denotes the propagation coefficients between the receive antenna array and the first receive metasurface layer. These coefficients are also modeled using (\ref{eq: Rayleigh-Sommerfield}).
	
	The wave-based precoding at the TX-SIM and combining at the RX-SIM are expressed as follows
	\begin{align}
		\mathbf{V}_{\TX}= & \boldsymbol{\Theta}_{\TX}^{L}\mathbf{\Omega}_{\TX}^{L-1}\boldsymbol{\Theta}_{\TX}^{L-1}\mathbf{\Omega}_{\TX}^{L}\cdots\boldsymbol{\Theta}_{\TX}^{2}\mathbf{\Omega}_{\TX}^{2}\boldsymbol{\Theta}_{\TX}^{1}\mathbf{\Omega}_{\TX}^{1}\in\mathbb{C}^{N\times S},\label{eq:gen_A}\\
		\mathbf{V}_{\RX}= & \mathbf{\Omega}_{\RX}^{1}\boldsymbol{\Theta}_{\RX}^{1}\mathbf{\Omega}_{\RX}^{2}\boldsymbol{\Theta}_{\RX}^{2}\cdots\mathbf{\Omega}_{\RX}^{K-1}\boldsymbol{\Theta}_{\RX}^{K-1}\mathbf{\Omega}_{\RX}^{K}\boldsymbol{\Theta}_{\RX}^{K}\in\mathbb{C}^{S\times M},\label{eq:gen_B}
	\end{align}
	where $\boldsymbol{\Theta}_{\TX}^{l}=\diag(\boldsymbol{\theta}_{\TX}^{l})\in\mathbb{C}^{N\times N}$ and $\boldsymbol{\Theta}_{\RX}^{k}=\diag(\boldsymbol{\theta}_{\RX}^{k})\in\mathbb{C}^{M\times M}$. Let $\tilde{\mathbf{H}}\in\mathbb{C}^{M\times N}$ represent the channel between the TX-SIM and RX-SIM. Then, the effective channel between the transmit and receive antennas is given by
	\begin{equation}
		\mathbf{H}=\mathbf{V}_{\RX}\tilde{\mathbf{H}}\mathbf{V}_{\TX}\in\mathbb{C}^{S\times S}.\label{eq:effective_channel}
	\end{equation}
	In SIM-aided HMIMO systems, each receive antenna $s$ is designed to capture the signal transmitted from its corresponding transmit antenna $s$, while signals from other transmit antennas $(j\neq s)$ are treated as interference. As a result, the AR for the SIM-aided HMIMO system is given by
	\begin{equation}
		R=\sum\nolimits_{s=1}^{S}\log_{2}\Bigl(1+\frac{|[\mathbf{H}]_{s,s}|^{2}p_{s}}{\sum\nolimits_{j\neq s}^{S}|[\mathbf{H}]_{s,j}|^{2}p_{j}+\sigma^{2}}\Bigr),\label{eq:SINRs}
	\end{equation}
	where $[\mathbf{H}]_{s,j}$ is the effective channel between the $j$-th transmit antenna and the $s$-th receive antenna, $p_{s}$ is the power allocated to the $s$-th transmit antenna, and $\sigma^{2}$ is the noise power.
\vspace{-8pt}
\subsection{Problem Formulations}	
In this paper, we consider two problem formulations that aim to maximize the AR by jointly optimizing the SIM phase shifts and power allocation. The first one arises from direct RMax in (\ref{eq:SINRs}), stated as 
	\begin{subequations}	
		\begin{IEEEeqnarray}{rCl}
		&\underset{_{\mathbf{p},\boldsymbol{\theta}_{\TX},\boldsymbol{\theta}_{\RX}}}{\max}  &\quad R(\mathbf{p},\boldsymbol{\theta}_{\TX},\boldsymbol{\theta}_{\RX}),
		\\*[-0.1\normalbaselineskip]
		&{\rm s.t.}  & |\theta_{\TX,n}^{l}|=1,\thinspace\forall n\in\mathcal{N},\forall l\in\mathcal{L}, \label{cnt:PHI}
		\\*[-0.1\normalbaselineskip]
		({\mathcal{P}_1}) : \smash{\left\{
			\IEEEstrut[10\jot]
			\right.}
		&  &|\theta_{\RX,m}^{k}|=1,\thinspace\forall m\in\mathcal{M},\forall k\in\mathcal{K}, \label{cnt:PSI}
		\\*[-0.1\normalbaselineskip]
		&  & \sum\nolimits_{s=1}^{S}p_{s}=P_{t}\label{eq:SPC}.
	\end{IEEEeqnarray}
\end{subequations}
where $\mathbf{p}=[p_{1},\cdots,p_{S}]\trans$, $P_{t}$ is the total transmit power at the TX-SIM, $\boldsymbol{\theta}_{\TX}=[(\boldsymbol{\boldsymbol{\theta}}_{\TX}^{1})\trans,\cdots,(\boldsymbol{\boldsymbol{\theta}}_{\TX}^{L})\trans]\trans\in\mathbb{C}^{NL\times1}$, and $\boldsymbol{\boldsymbol{\theta}}_{\RX}=[(\boldsymbol{\boldsymbol{\theta}}_{\RX}^{1})\trans,\cdots,(\boldsymbol{\boldsymbol{\theta}}_{\RX}^{K})\trans]\trans\in\mathbb{C}^{MK\times1}$. Note that even for fixed phase shifts, $({\mathcal{P}_1})$ is non-convex and indeed NP-hard, due to the inter-stream interference.
\par The second formulation is based on the IMin, given by
		\begin{equation}\nonumber
		(\mathcal{P}_{2}):\Bigl\{\begin{array}{rl}
			\underset{\boldsymbol{\theta}_{\TX},\boldsymbol{\theta}_{\RX}}{\min} & \sum_{s=1}^{S}\sum\nolimits_{j\neq s}^{S}|[\mathbf{H}]_{s,j}|^{2}\triangleq\left\Vert \mathbf{L}\vect(\mathbf{H})\right\Vert ^{2},\\
			\st & (\ref{cnt:PHI}) \quad\rm{and} \quad (\ref{cnt:PSI}).
		\end{array}
	\end{equation}
where $\mathbf{L}\in\mathbb{Z}^{S(S-1)\times S^{2}}$ is the matrix extracting the off-diagonal elements of the square matrix $\mathbf{H}$. In other words, $(\mathcal{P}_{2})$ aims to diagonalize the effective channel $\mathbf{H}$ in (\ref{eq:effective_channel}), which is inspired by the zero-forcing method. Note that $(\mathcal{P}_{2})$ does not include power allocation optimization. This formulation is motivated by two key advantages. First, as shown later, it admits an efficient iterative algorithm based on closed-form expression. Second, in ideal cases, the inter-stream interference would be completely canceled. Hence, the power allocation can be found efficiently by WF algorithm. The solutions to $({\mathcal{P}_1})$ and $(\mathcal{P}_{2})$ are presented in the following sections.	
\section{Proposed Solution to $({\mathcal{P}_1})$\label{sec:PropSolution}}
We adopt the AO-based approach to solve $(\mathcal{P}_{1})$, leading to SIM phase shifts design and power allocation subproblems.	

\textit{Phase-shift optimization}: With $\mathbf{p}$ fixed in $(\mathcal{P}_{1})$, the optimization problem for $\{\boldsymbol{\theta}_{\TX},\boldsymbol{\theta}_{\RX}\}$ is given by
    \begin{equation}
    (\mathcal{P}_{3}) \triangleq \{\underset{_{\boldsymbol{\theta}_{\TX},\boldsymbol{\theta}_{\RX}}}{\max}\ R(\boldsymbol{\theta}_{\TX},\boldsymbol{\theta}_{\RX})\ |\ (\ref{cnt:PHI}) \ \textrm{and} \ (\ref{cnt:PSI})\}
	\end{equation}
	However, $(\mathcal{P}_{3})$ is still intractable due to the nonconvexity of $R(\boldsymbol{\theta}_{\TX},\boldsymbol{\theta}_{\RX})$ and the unit-modulus constraints of $\boldsymbol{\theta}_{\TX}$ and $\boldsymbol{\theta}_{\RX}$.
	Considering the fact that the constraints for $\boldsymbol{\theta}_{\TX}$ and $\boldsymbol{\theta}_{\RX}$ are decoupled, we optimize one variable at a time while holding the other fixed. Moreover, the unit modulus constraints of $\boldsymbol{\theta}_{\TX}$ and $\boldsymbol{\theta}_{\RX}$ define the Riemannian manifold, which motivates the use of an RMO-based method to optimize $\{\boldsymbol{\theta}_{\TX},\boldsymbol{\theta}_{\RX}\}$. Specifically, we employ the Riemannian Broyden-Fletcher-Goldfarb-Shanno (BFGS) algorithm, as detailed in\cite{absil2009optimization}. Note that other RMO-based methods, such as the Conjugate-Gradient, Barzilai-Borwein, Trust-Region methods, or the PG-based method \cite{absil2009optimization}
	can also be employed to address $(\mathcal{P}_{3})$. The reason we choose the BFGS is that it is a well-known quasi-Newton method that has shown to be one of the most effective algorithms for solving non-convex problems \cite{absil2009optimization}. At the core of the RMO is the calculation of the Euclidean gradients (EGs) of $R(\boldsymbol{\theta}_{\TX},\boldsymbol{\theta}_{\RX})$ w.r.t $\boldsymbol{\theta}_{\TX}^{\ast}$ and $\boldsymbol{\theta}_{\RX}^{\ast}$. To this end, the EGs of the $R(\boldsymbol{\theta}_{\TX},\boldsymbol{\theta}_{\RX})$ w.r.t $\boldsymbol{\theta}_{\TX}^{\ast}$ and w.r.t $\boldsymbol{\theta}_{\RX}^{\ast}$ are respectively given by
	\begin{equation}
		\resizebox{.9\hsize}{!}{\ensuremath{\nabla_{\boldsymbol{\theta}_{\TX}}R(\boldsymbol{\theta}_{\TX},\boldsymbol{\theta}_{\RX})=\left[(\nabla_{\boldsymbol{\theta}_{\TX}^{1}}R(\boldsymbol{\theta}_{\TX},\boldsymbol{\theta}_{\RX}))\trans,\cdots,(\nabla_{\boldsymbol{\theta}_{\TX}^{L}}R(\boldsymbol{\theta}_{\TX},\boldsymbol{\theta}_{\RX}))\trans\right]\trans}},\label{eq:euclidean_gradient_TX_SIM}
	\end{equation}
	\begin{equation}
		\resizebox{.9\hsize}{!}{\ensuremath{\nabla_{\boldsymbol{\theta}_{\RX}}R(\boldsymbol{\theta}_{\TX},\boldsymbol{\theta}_{\RX})=\left[(\nabla_{\boldsymbol{\theta}_{\RX}^{1}}R(\boldsymbol{\theta}_{\TX},\boldsymbol{\theta}_{\RX}))\trans,\cdots,(\nabla_{\boldsymbol{\theta}_{\RX}^{K}}R(\boldsymbol{\theta}_{\TX},\boldsymbol{\theta}_{\RX}))\trans\right]\trans}},\label{eq:euclidean_gradient_RX_SIM}
	\end{equation}
	where $\nabla_{\boldsymbol{\theta}_{\TX}^{l}}R(\boldsymbol{\theta}_{\TX},\boldsymbol{\theta}_{\RX})$ and $\nabla_{\boldsymbol{\theta}_{\RX}^{k}}R(\boldsymbol{\theta}_{\TX},\boldsymbol{\theta}_{\RX})$ are in \thmref{grad:theta}.
	\begin{thm}
		\label{thm:grad:theta}A closed-form expression for $\nabla_{\boldsymbol{\theta}_{\TX}^{l}}R(\boldsymbol{\theta}_{\TX},\boldsymbol{\theta}_{\RX})$ and $\nabla_{\boldsymbol{\theta}_{\RX}^{k}}R(\boldsymbol{\theta}_{\TX},\boldsymbol{\theta}_{\RX})$
		are respectively given by (\ref{eq:phi_derivative}) and (\ref{eq:psi_derivative}), shown at the top of the following page, where $\mathbf{V}_{\TX}^{1-}=\mathbf{I}_{N},$ $\mathbf{V}_{\TX}^{L+}=\mathbf{I}_{N}$,$\mathbf{V}_{\RX}^{1-}=\mathbf{I}_{M},$
		$\mathbf{V}_{\RX}^{K+}=\mathbf{I}_{M}$,
		\[
		\begin{array}{rl}
			\mathbf{V}_{\TX}^{l-}= & \mathbf{\Omega}_{\TX}^{l}\boldsymbol{\Theta}_{\TX}^{l-1}\mathbf{\Omega}_{\TX}^{l-1}\cdots\boldsymbol{\Theta}_{\TX}^{2}\mathbf{\Omega}_{\TX}^{2}\boldsymbol{\Theta}_{\TX}^{1},2\leq l\leq L,\\
			\mathbf{V}_{\TX}^{l+}= & \boldsymbol{\Theta}_{\TX}^{L}\mathbf{\Omega}_{\TX}^{L}\cdots\boldsymbol{\Theta}_{\TX}^{l+1}\mathbf{\Omega}_{\TX}^{l+1},1\leq l\leq L-1,\\
			\mathbf{V}_{\RX}^{k-}= & \boldsymbol{\Theta}_{\RX}^{1}\mathbf{\Omega}_{\RX}^{2}\boldsymbol{\Theta}_{\RX}^{2}\cdots\mathbf{\Omega}_{\RX}^{k-1}\boldsymbol{\Theta}_{\RX}^{k-1}\mathbf{\Omega}_{\RX}^{k},2\leq k\leq K,\thinspace{\rm \mathit{and}}\\
			\mathbf{V}_{\RX}^{k+}= & \mathbf{\Omega}_{\RX}^{k+1}\boldsymbol{\Theta}_{\RX}^{k+1}\cdots\mathbf{\Omega}_{\RX}^{K}\boldsymbol{\Theta}_{\RX}^{K},1\leq k\leq K-1.
		\end{array}
		\]
		\begin{figure*}
			\begin{equation}
				\resizebox{.948\hsize}{!}{\ensuremath{\nabla_{\boldsymbol{\theta}^{l}_{\TX}}R(\boldsymbol{\theta}_{\TX},\boldsymbol{\theta}_{\RX})=\frac{1}{\ln(2)}\sum_{s=1}^{S}\Bigl(\frac{\sum\nolimits_{j=1}^{S}[\mathbf{H}]_{s,j}p_{j}\ensuremath{\vecd(\mathbf{V}_{\TX}^{l-}[\mathbf{\Omega}_{\TX}^{1}]_{:,j}[\mathbf{V}_{\RX}]_{s,:}\tilde{\mathbf{H}}\mathbf{V}_{\TX}^{l+})^{\ast}}}{\sum\nolimits_{j}^{S}|[\mathbf{H}]_{s,j}|^{2}p_{j}+\sigma^{2}}-\frac{\sum\nolimits_{j\neq s}^{S}[\mathbf{H}]_{s,j}p_{j}\ensuremath{\vecd(\mathbf{V}_{\TX}^{l-}[\mathbf{\Omega}_{\TX}^{1}]_{:,j}[\mathbf{V}_{\RX}]_{s,:}\tilde{\mathbf{H}}\mathbf{V}_{\TX}^{l+})^{\ast}}}{\sum\nolimits_{j\neq s}^{S}|[\mathbf{H}]_{s,j}|^{2}p_{j}+\sigma^{2}}\Bigr)}}\label{eq:phi_derivative}
			\end{equation}
			
			\begin{equation}
				\resizebox{.948\hsize}{!}{\ensuremath{\nabla_{\boldsymbol{\theta}^{k}_{\RX}}R(\boldsymbol{\theta}_{\TX},\boldsymbol{\theta}_{\RX})=\frac{1}{\ln(2)}\sum_{s=1}^{S}\Bigl(\frac{\sum\nolimits_{j=1}^{S}[\mathbf{H}]_{s,j}p_{j}\vecd(\mathbf{V}_{\RX}^{k+}\tilde{\mathbf{H}}[\mathbf{V}_{\TX}]_{:,j}[\mathbf{\Omega}_{\RX}^{1}]_{s,:}\mathbf{V}_{\RX}^{k-})^{\ast}}{\sum\nolimits_{j}^{S}|[\mathbf{H}]_{s,j}|^{2}p_{j}+\sigma^{2}}-\frac{\sum\nolimits_{j\neq s}^{S}[\mathbf{H}]_{s,j}p_{j}\vecd(\mathbf{V}_{\RX}^{k+}\tilde{\mathbf{H}}[\mathbf{V}_{\TX}]_{:,j}[\mathbf{\Omega}_{\RX}^{1}]_{s,:}\mathbf{V}_{\RX}^{k-})^{\ast}}{\sum\nolimits_{j\neq s}^{S}|[\mathbf{H}]_{s,j}|^{2}p_{j}+\sigma^{2}}\Bigr)}}\label{eq:psi_derivative}
			\end{equation}
			
			\hrulefill{}
		\end{figure*}
	\end{thm}
	\begin{IEEEproof}
		See the Appendix.
	\end{IEEEproof}
	\textit{Power allocation subproblem}: Next, after optimizing $\{\boldsymbol{\theta}_{\TX},\boldsymbol{\theta}_{\RX}\}$, we optimize $\mathbf{p}$ by fixing $\{\boldsymbol{\theta}_{\TX},\boldsymbol{\theta}_{\RX}\}$ in $(\mathcal{P}_{1})$, which leads to the following power allocation problem:   
    \begin{equation}
    (\mathcal{P}_{4}) \triangleq \{\underset{_{\mathbf{p}}}{\max}\ R(\mathbf{p})\ |\  (\ref{eq:SPC}) \}
	\end{equation}
	Problem $(\mathcal{P}_{4})$ is a classical power allocation problem. To solve it, we adopt the prevailing weighted minimum mean square error (WMMSE) method as detailed in \cite{shi2011iteratively}.
    
    The overall AO procedure to solve $({\mathcal{P}_1})$ is summarized in \algref{Alg1}. Specifically, the initial $\{\boldsymbol{\theta}_{\TX}^{(0)},\boldsymbol{\theta}_{\RX}^{(0)}\}$ are randomly generated and $\mathbf{p}^{(0)}$ is initialized such that $p_{1}=\cdots=p_{S}$.
	\begin{algorithm}[b]
		\small 
		\SetAlgoLined
		\DontPrintSemicolon
		\SetKwRepeat{Do}{do}{while}
		\SetKwInput{Initialize}{Initialize}
		\SetKwInOut{Input}{Input}
		\SetKwInOut{Output}{Output}\Initialize{$\boldsymbol{\theta}_{\TX}^{(0)}$, $\boldsymbol{\theta}_{\RX}^{(0)}$, $\mathbf{p}^{(0)}$, and $q=0$.}
		
		\Repeat{convergence }{
			
			Fix $\boldsymbol{\theta}_{\RX}$ in $(\mathcal{P}_{3})$, obtain $\boldsymbol{\theta}_{\TX}^{(q+1)}$ via the RMO using the EG given in (\ref{eq:euclidean_gradient_TX_SIM}) and (\ref{eq:phi_derivative})\;
			
			Fix $\boldsymbol{\theta}_{\TX}$ in $(\mathcal{P}_{3})$, obtain $\boldsymbol{\theta}_{\RX}^{(q+1)}$ via the RMO using the EG given in (\ref{eq:euclidean_gradient_RX_SIM}) and (\ref{eq:psi_derivative})\;
			
			Find $\mathbf{p}^{(q+1)}$ by solving $(\mathcal{P}_{4})$ via the WMMSE method\;
			
			$q \leftarrow q+1$ \;
			
		}\caption{Proposed algorithm for solving $({\mathcal{P}_1})$\label{alg:Alg1}}
	\end{algorithm}
	
	\emph{Complexity analysis}: The complexity of \algref{Alg1} mainly relies on the calculation of the EGs w.r.t $\boldsymbol{\theta}_{\TX}$ and $\boldsymbol{\theta}_{\RX}$. These calculations have complexities of $\mathcal{O}(L^{2}N^{3}+LN^{2}S^{2}+LNMS)$ and $\mathcal{O}(K^{2}M^{3}+LM^{2}S^{2}+LNMS)$, respectively. Another significant contribution to the overall complexity comes from the computation of $\mathbf{V}_{\TX}$ and $\mathbf{V}_{\RX}$, with complexities of $\mathcal{O}(LN^{3})$ and $\mathcal{O}(KM^{3})$, respectively. Hence, the total complexity of \algref{Alg1} is $\mathcal{O}\bigl(I_{O}\bigl(I_{\boldsymbol{\theta}_{\TX}}\bigl(L^{2}N^{3}+LN^{2}S^{2}+LNMS+KM^{3}\bigr)+I_{\boldsymbol{\theta}_{\RX}}\bigl(K^{2}M^{3}+LM^{2}S^{2}+LNMS+LN^{3}\bigr)\bigr)\bigr)$,
	where $I_{\boldsymbol{\theta}_{\TX}}$ and $I_{\boldsymbol{\theta}_{\RX}}$ denote the number of inner RMO iterations for updating $\boldsymbol{\theta}_{\TX}$ and $\boldsymbol{\theta}_{\RX}$, respectively, and $I_{O}$ represents the number of outer loop iterations.
	
	\section{Proposed Solution to $(\mathcal{P}_{2})$}
	
	In this section, we propose an iterative method to solve $(\mathcal{P}_{2})$. Leveraging its underlying structure, our strategy is to sequentially optimize the phase shift of each meta-atom, allowing us to achieve a closed-form solution. To proceed, consider the $l$-th metasurface layer of the TX-SIM, we rewrite $(\mathcal{P}_{2})$ as a function of $\{\theta_{\TX,n}^{l}\}n\in\mathcal{N}$ as
	\begin{subequations}
		\begin{align}
			&\bigl\Vert\mathbf{L}\vect(\mathbf{V}_{\RX}\tilde{\mathbf{H}}\mathbf{V}_{\TX}^{l+}\boldsymbol{\Theta}_{\TX}^{l}\mathbf{V}_{\TX}^{l-}\mathbf{\Omega}_{\TX}^{1})\bigr\Vert^{2}\label{eq:g11}\\
			& =\bigl\Vert\mathbf{L}\bigl(\bigl(\mathbf{V}_{\TX}^{l-}\mathbf{\Omega}_{\TX}^{1}\bigr)\trans\otimes\bigl(\mathbf{V}_{\RX}\tilde{\mathbf{H}}\mathbf{V}_{\TX}^{l+}\bigr)\bigr)\vect\bigl(\boldsymbol{\Theta}_{\TX}^{l}\bigr)\bigr\Vert^{2}\label{eq:g1}\\
			& =\bigl\Vert\mathbf{L}\bigl(\bigl(\mathbf{V}_{\TX}^{l-}\mathbf{\Omega}_{\TX}^{1}\bigr)\trans\otimes\bigl(\mathbf{V}_{\RX}\tilde{\mathbf{H}}\mathbf{V}_{\TX}^{l+}\bigr)\bigr)\tilde{\mathbf{L}}_{\TX}\boldsymbol{\theta}_{\TX}^{l}\bigr\Vert^{2}\label{eq:g2}\\
			& =\bigl\Vert\mathbf{E}_{\TX}^{l}\boldsymbol{\theta}_{\TX}^{l}\bigr\Vert^{2}\overset{(a)}{=}\Tr\bigl\{\bigl(\sum\nolimits_{i\neq n}^{N}\mathbf{e}_{\TX,i}^{l}\theta_{\TX,i}^{l}\bigr)\herm\bigl(\sum\nolimits_{i\neq n}^{N}\mathbf{e}_{\TX,i}^{l}\theta_{\TX,i}^{l}\bigr)\nonumber \\
			& +2\mathfrak{R}\bigl\{\theta_{n}^{l\ast}\bigl(\mathbf{e}_{\TX,n}^{l}\bigr)\herm\bigl(\sum\nolimits_{i\neq n}^{N}\mathbf{e}_{\TX,i}^{l}\theta_{\TX,i}^{l}\bigr)\bigr\}+\bigl(\mathbf{e}_{\TX,n}^{l}\bigr)\herm\mathbf{e}_{\TX,n}^{l}\bigr\},\label{eq:g3}
		\end{align}
	\end{subequations}
	where (\ref{eq:g1}) is due to the identity $\vect(\mathbf{A}\mathbf{B}\mathbf{C})=(\mathbf{C}\trans\otimes\mathbf{A})\vect(\mathbf{B})$, $\tilde{\mathbf{L}}_{\TX}\in\mathbb{Z}^{N^{2}\times N}$ maps the diagonal elements of a matrix $\mathbf{X}\in\mathbb{C}^{N\times N}$ into $\vect(\mathbf{X})$, $\mathbf{E}_{\TX}^{l}=\mathbf{L}\bigl((\mathbf{V}_{\TX}^{l-}\mathbf{\Omega}_{\TX}^{1})\trans\otimes(\mathbf{V}_{\RX}\tilde{\mathbf{H}}\mathbf{V}_{\TX}^{l+})\bigr)\tilde{\mathbf{L}}_{\TX}\in\mathbb{C}^{S(S-1)\times N}$, and $\mathbf{e}_{\TX,i}^{l}\in\mathbb{C}^{S(S-1)\times1},$ is the $i$-th column vector of $\mathbf{E}_{\TX}^{l}$. Now, for fixed $\{\theta_{\TX,i}^{l}\}\ \forall i\neq n\in\mathcal{N},\ \forall l\in\mathcal{L}$ and $\boldsymbol{\theta}_{\RX}$, it is straightforward to see that the optimal value of $\theta_{\TX,n}^{l}$ for $(\mathcal{P}_{2})$, is found as 
	\begin{equation}
		\theta_{\TX,n}^{l}=\exp\bigl(j\arg\{-\bigl(\mathbf{e}_{\TX,n}^{l}\bigr)\herm\bigl(\sum\nolimits_{i\neq n}^{N}\mathbf{e}_{\TX,i}^{l}\theta_{\TX,i}^{l}\bigr)\}\bigr).\label{eq:closed_form_TX}
	\end{equation}
	The same approach is applied to optimize $\{\theta_{\RX,m}^{k}\}\ m\in\mathcal{M}$ for the RX-SIM, with other meta-atoms held fixed. Specifically, $g$ is expressed in terms of $\{\theta_{\RX,m}^{k}\}\ m\in\mathcal{M}$, following similar steps in (\ref{eq:g11}) through (\ref{eq:g3}). This results in the optimal $\theta_{\RX,m}^{k}$, for fixed $\{\theta_{\RX,i}^{k}\}\ \forall i\neq m\in\mathcal{M},\ \forall k\in\mathcal{K}$ and $\boldsymbol{\theta}_{\TX}$, given by 
	\begin{equation}
		\theta_{\RX,m}^{k}=\exp\bigl(j\arg\{-\bigl(\mathbf{e}_{\RX,m}^{k}\bigr)\herm\bigl(\sum\nolimits_{i\neq m}^{M}\mathbf{e}_{\RX,i}^{k}\theta_{\RX,i}^{k}\bigr)\}\bigr),\label{eq:closed_form_RX}
	\end{equation}
	where $\mathbf{e}_{\RX,i}^{k}\in\mathbb{C}^{S(S-1)\times1}$ is the $i$-th column vector of $\mathbf{E}_{\RX}^{k}=\mathbf{L}(\mathbf{V}_{\RX}^{k+}\tilde{\mathbf{H}}\mathbf{V}_{\TX})\trans\otimes(\mathbf{\Omega}_{\RX}^{1}\mathbf{V}_{\RX}^{k-})\tilde{\mathbf{L}}_{M}\in\mathbb{C}^{S(S-1)\times M}$.
	
	The iterative process keeps updating $\theta_{\TX,n}^{l}$ and $\theta_{\RX,m}^{k}$ according to (\ref{eq:closed_form_TX}) and (\ref{eq:closed_form_RX}), respectively, until convergence. After the optimization of $\boldsymbol{\theta}_{\TX}$ and $\boldsymbol{\theta}_{\RX}$ is completed, we ignore the interference term (i.e. setting $\sum\nolimits_{j\neq s}^{S}|[\mathbf{H}]_{s,j}|^{2}=0$ in $(\mathcal{P}_{4})$ and then applying the WF algorithm to find the optimal power allocation $\mathbf{p}$. The overall procedures are summarized in \algref{Alg2}.
	\begin{algorithm}[t]
		\small 
		\SetAlgoLined
		\DontPrintSemicolon
		\SetKwRepeat{Do}{do}{while}
		\SetKwInput{Initialize}{Initialize}
		\SetKwInOut{Input}{Input}
		\SetKwInOut{Output}{Output}
		
		\Initialize{$\boldsymbol{\theta}_{\TX}^{(0)}$, $\boldsymbol{\theta}_{\RX}^{(0)}$, and $q=0$.}		
		\BlankLine		
		\Repeat{convergence }{			
		      Compute $\boldsymbol{\theta}_{\TX}^{(q+1)}$ using (\ref{eq:closed_form_TX})\;
			
			Compute $\boldsymbol{\theta}_{\RX}^{(q+1)}$ using (\ref{eq:closed_form_RX})\;			
			$q=q+1$\;			
		}		
		Set  $\sum\nolimits_{j\neq s}^{S}|[\mathbf{H}]_{s,j}|^{2}=0$ and apply the WF to $(\mathcal{P}_{4})$ to find $\mathbf{p}$.
		
		\caption{Proposed algorithm for solving $(\mathcal{P}_{2})$ \label{alg:Alg2}}
	\end{algorithm}
	
    \emph{Complexity analysis}: The computational complexity of \algref{Alg2} mainly depends on calculations of $\mathbf{E}_{\TX}^{l}\ \forall l\in\mathcal{L}$ and $\mathbf{E}_{\RX}^{k}\ \forall k\in\mathcal{K}$. These calculations have complexities of $\mathcal{O}(L^{2}N^{3}+LN^{2}S^{2}+LNMS)$ and $\mathcal{O}(K^{2}M^{3}+KM^{2}S^{2}+KNMS)$, respectively. Consequently, the total complexity of \algref{Alg2} is $\mathcal{O}\big(I_{O}\bigl(L^{2}N^{3}+LN^{2}S^{2}+LNMS+LN^{3}	+K^{2}M^{3}+KM^{2}S^{2}+KNMS+KM^{3}\bigr)\big)$.
\section{Proposed Hybrid Algorithm}	
The consideration of two formulations in our work deserves further discussion, as it leads to the development of the proposed hybrid approach, which effectively scales the AR with the number of SIM layers. Extensive numerical experiments indicate that \algref{Alg1} is highly sensitive to the initial phase shifts. Specifically, when initialized randomly, the performance of \algref{Alg1} degrades as the number of SIM layers exceeds a certain threshold. While this phenomenon was also reported in previous studies, \algref{Alg2} exhibits a consistent increase in AR as the number of SIM layers grows, although performing worse than \algref{Alg1} for small to moderate numbers of SIM layers. These observations suggest that \algref{Alg1} struggles to escape ``deep valley'' when the problem size becomes large. To overcome this issue, we propose a hybrid method that leverages the strengths of both algorithms presented in preceding sections. In this hybrid approach, the phase shifts returned by \algref{Alg2} are used to initialize \algref{Alg1}. This initialization strategy follows an intuitive heuristic approach, which requires an analytical proof that is beyond the scope of this paper. As shown in the next section, this simple method yields improved ARs across the entire range of SIM layers.
	
	\section{Simulation Results\label{sec:SimResults}}
	
	In this section, we present numerical simulation results to evaluate and compare the performance of our proposed algorithms against two benchmarks: the existing PG-based method \cite{an2023stacked} and a fully digital precoding MIMO scheme. The simulation settings are taken from \cite{papazafeiropoulos2024achievable,an2023stacked}, as follows: the system operates at a carrier frequency of $f_{c}=6\thinspace{\rm GHz}$, corresponding to a wavelength of $\lambda=50\thinspace{\rm mm}$. The transmit and receive antennas are modeled as uniform linear arrays, parallel to the $x$-axis, with their midpoints positioned at $(0,0,0)$ and $(0,0,d)$, respectively. The SIM layers are modeled as uniform planar arrays, aligned parallel to the $xy$-plane, with their centers positioned along the $z$-axis. Both the TX-SIM and RX-SIM are assumed to have equal thicknesses, $D_{\TX}=D_{\RX}=0.1\thinspace{\rm m}$, and the corresponding spacing between adjacent metasurface layers are $\nicefrac{D_{\TX}}{L}$ and $\nicefrac{D_{\RX}}{K}$, respectively. The meta-atom spacing is set to $\nicefrac{\lambda}{2}$ and the surface area of each meta-atom is $A=\nicefrac{\lambda^{2}}{4}.$ Other system parameters are set as $P_{\TX}=20\,{\rm dBm}$, $\sigma^{2}=-110\,{\rm dBm}$, $S=4$, $N=M=100$, and $L=K=7$, unless otherwise specified.
	
	The spatially-correlated HMIMO channel model is adopted to generate the channel. Specifically, the channel matrix between TX-SIM and RX-SIM is modeled as $\tilde{\mathbf{H}}=\mathbf{R}_{\RX}^{1/2}\mathbf{\tilde{\mathbf{H}}}_{w}\mathbf{R}_{\TX}^{1/2},$
	where $\mathbf{\tilde{\mathbf{H}}}_{w}$ is an $N\times M$ matrix composed of independent and identically distributed complex Gaussian random variables, i.e. $\mathbf{\tilde{\mathbf{H}}}_{w}\thicksim\mathcal{CN}(\boldsymbol{0},\xi\mathbf{I}_{N}\otimes\mathbf{I}_{M})$, where $\xi$ presents the distance-dependent path loss between the transceivers and is given by $\xi[{\rm dB}]=10a_{1}\log_{10}(\nicefrac{4\pi d_{0}}{\lambda})+10a_{2}\log_{10}(\nicefrac{d}{d_{0}}),\thinspace d\geq d_{0}$, where $d_{0}$ and $d$ are the reference and link distances in meters, respectively, and $a_{1}$ and $a_{2}$ are constants. The matrices $\mathbf{R}_{\TX}\in\mathbb{C}^{N\times N}$ and $\mathbf{R}_{\RX}\in\mathbb{C}^{M\times M}$ denote the spatial correlation matrices at the TX-SIM and RX-SIM, respectively, with their entries defined in \cite{an2023stacked}. For the simulations in this paper, we set $d_{0}=1\,{\rm m}$, $d=240\,{\rm m}$, $a_{1}=2$, and $a_{2}=3.5$. The simulation results are obtained by averaging over $100$ independent experiments.
	\begin{figure}[b]
		\begin{centering}
			\includegraphics[scale=1]{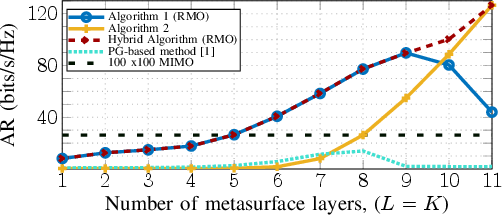}
			\par\end{centering}
		\caption{\label{fig:Rate_vs_LK}AR versus the number of metasurface layers ($L=K$).}
	\end{figure}	
	In \figref{Rate_vs_LK}, we present the AR of different numbers of metasurface layers for a fixed SIM thickness. The results highlight that the AR increases with the number of SIM layers, with the proposed algorithms outperforming the PG-based method \cite{an2023stacked} and the conventional digital precoding approach for the massive MIMO counterpart. However, as mentioned earlier, the performance of \algref{Alg1} deteriorates as the number of SIM layers becomes large, a trend also seen in the PG-based method \cite{an2023stacked}. In contrast, the performance of \algref{Alg2} improves significantly for large numbers of SIM layers. Most notably, the \textbf{Hybrid Algorithm} consistently achieves superior performance across the entire range of SIM layers. This hybrid approach offers a steadily increasing AR as the number of SIM layers grows, contradicting earlier findings in \cite{an2023stacked} and \cite{papazafeiropoulos2024achievable}, which suggested that the AR would either decrease or saturate with an excessive number of metasurface layers for a fixed SIM thickness.	
    \begin{figure}[t]
		\begin{centering}
			\includegraphics[scale=1]{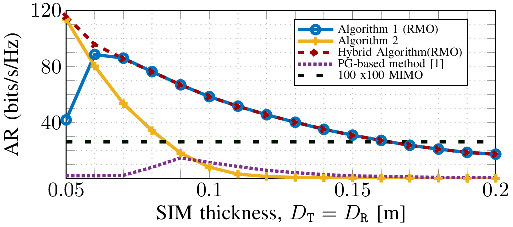}
			\par\end{centering}
		\caption{\label{fig:Rate_vs_thickness}AR versus the SIM thickness, $D_{\TX}=D_{\RX}\,[{\rm m}]$}
	 \end{figure}
    \begin{figure}[b]
        \begin{centering}
            \includegraphics[scale=1]{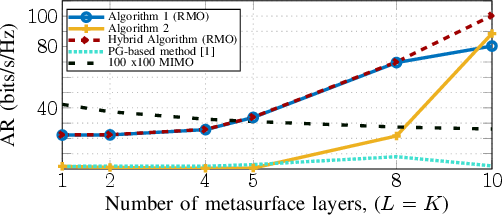}
            \par\end{centering}
        \caption{\label{fig:Rate_vs_layers_Nfixed}AR versus the number of SIM layers with a fixed total of $1000$ meta-atoms at the TX-SIM and RX-SIM.}
    \end{figure}	
    \par In \figref{Rate_vs_thickness}, we show the AR of SIM-aided HMIMO systems for different SIM thicknesses, with the number of metasurface layers fixed at $L=K=7$. \figref{Rate_vs_thickness} reveals that the AR decreases as the SIM thickness increases. Similar to \figref{Rate_vs_LK}, we observe that for a very small SIM thickness or densely packed metasurface layers, the AR of \algref{Alg2} and the \textbf{Hybrid Algorithm} significantly improves, which contrasts with the findings reported in \cite{an2023stacked} and \cite{papazafeiropoulos2024achievable}.
    
    In \figref{Rate_vs_layers_Nfixed}, we evaluate the AR performance with  $1000$ meta-atoms fixed in both the TX-SIM and RX-SIM, while varying the number of SIM layers such that $N={1000}/{L}$ and $M={1000}/{K}$. We observe that the AR increases slightly for small values of ${L}$ and ${K}$, but increases sharply when  $L=K>4$, which demonstrates that increasing the number of layers under a fixed SIM thickness significantly enhances AR performance.
	
  \section{Conclusion and Future work\label{sec:Conclusion}}
 In this paper, we have investigated the AR of SIM-aided HMIMO systems by jointly optimizing SIM phase shifts and power allocation. Specifically, we have introduced two problem formulations to maximize the AR. Numerical experiments have shown that the proposed algorithms significantly outperform existing benchmarks. Most importantly, unlike prior studies, we have demonstrated that increasing the number of metasurface layers while maintaining a fixed SIM thickness results in notable AR improvements, provided the propagation between metasurfaces adheres to Rayleigh-Sommerfield diffraction theory. While our proposed methods yield significant performance gains, they do not guarantee a globally optimal solution. For future work, it would be interesting to perform an analytical analysis of the proposed hybrid optimization approach and evaluate its performance under different inter-layer channel models.
	
	\appendix[Proof of \thmref{grad:theta}]{}
	
	The gradient of $R(\boldsymbol{\theta}_{\TX},\boldsymbol{\theta}_{\RX})$ w.r.t $\boldsymbol{\theta}_{\TX}^{l\ast}$ is given by
	\begin{align}
		\ensuremath{\nabla_{\boldsymbol{\theta}_{\TX}^{l}}R(\boldsymbol{\theta}_{\TX},\boldsymbol{\theta}_{\RX})} & =\frac{1}{\ln(2)}\sum_{s=1}^{S}\Bigl(\frac{\sum_{j=1}^{S}[\mathbf{H}]_{s,j}\nabla_{\boldsymbol{\theta}_{\TX}^{l}}[\mathbf{H}]_{s,j}^{\ast}p_{j}}{\sum_{j=1}^{S}|[\mathbf{H}]_{s,j}|^{2}p_{j}+\text{\ensuremath{\sigma^{2}}}}\nonumber \\
		& \qquad-\frac{\sum_{j\neq s}^{S}[\mathbf{H}]_{s,j}\nabla_{\boldsymbol{\theta}_{\TX}^{l}}[\mathbf{H}]_{s,j}^{\ast}p_{j}}{\sum_{j\neq s}^{S}|[\mathbf{H}]_{s,j}|^{2}p_{j}+\sigma^{2}}\Bigr).\label{eq:gradFunc}
	\end{align}
	Next, consider the complex differential of $[\mathbf{H}]_{s,j}^{\ast}$ given by 
	\begin{subequations}
		\begin{align}
			d[\mathbf{H}]_{s,j}^{\ast} & =\ensuremath{\Tr\{([\mathbf{V}_{\RX}]_{s,:}\tilde{\mathbf{H}}\mathbf{V}_{\TX}^{l+}d\boldsymbol{\Theta}_{\TX}^{l}\mathbf{V}_{\TX}^{l-}[\mathbf{\Omega}_{\TX}^{1}]_{:,j})^{\ast}\}},\\
			& =\Tr\{(\mathbf{V}_{\TX}^{l-}[\mathbf{\Omega}_{\TX}^{1}]_{:,j}[\mathbf{V}_{\RX}]_{s,:}\tilde{\mathbf{H}}\mathbf{V}_{\TX}^{l+}d\boldsymbol{\Theta}_{\TX}^{l})^{\ast}\},\\
			& =\ensuremath{\vecd(\mathbf{V}_{\TX}^{l-}[\mathbf{\Omega}_{\TX}^{1}]_{:,j}[\mathbf{V}_{\RX}]_{s,:}\tilde{\mathbf{H}}\mathbf{V}_{\TX}^{l+})^{\ast}d\boldsymbol{\theta}_{\TX}^{\ast}}.\label{eq:eqdtheta}
		\end{align}
	\end{subequations}
	We have applied the trace property \cite[Eqn. (2.96)]{hjorungnes2011complex} and the matrix vectorization property \cite[Lemma 2.24]{hjorungnes2011complex} accordingly. Thus, using the results from \cite[Table 3.2]{hjorungnes2011complex},
	(\ref{eq:eqdtheta}) yields
	\begin{equation}
		\ensuremath{\nabla_{\boldsymbol{\theta}_{\TX}^{l}}[\mathbf{H}]_{s,j}^{\ast}=\vecd(\mathbf{V}_{\TX}^{l-}[\mathbf{\Omega}_{\TX}^{1}]_{:,j}[\mathbf{V}_{\RX}]_{s,:}\tilde{\mathbf{H}}\mathbf{V}_{\TX}^{l+})^{\ast}}.\label{eq:grad_TX-SIM}
	\end{equation}
	Substituting (\ref{eq:grad_TX-SIM}) into (\ref{eq:gradFunc}) gives (\ref{eq:phi_derivative}). Similar steps as above are used for $\nabla_{\boldsymbol{\theta}_{\RX}^{k}}[\mathbf{H}]_{s,j}^{\ast}$, thereby completing the proof.
	
	\bibliographystyle{IEEEtran}
	\bibliography{IEEEabrv,bahingayi_WCL2025-0968}
	
\end{document}